\documentclass[twocolumn,tighten,twocolappendix]{aastex63}

\usepackage{CJK}
\graphicspath{{./}{figures/}}

\submitjournal{ApJ}

\shorttitle{Forecasting Stellar Streams}
\shortauthors{Kamdar et al.}

\begin{document}
\begin{CJK*}{UTF8}{gbsn}

\title{Stellar Streams in the Galactic Disk: Predicted Lifetimes and Their Utility in Measuring the Galactic Potential}

\correspondingauthor{Harshil Kamdar}
\email{harshil.kamdar@cfa.harvard.edu}

\author[0000-0001-5625-5342]{Harshil Kamdar}
\affiliation{Center for Astrophysics $|$ Harvard \& Smithsonian, 60 Garden Street, Cambridge, MA 02138, USA}

\author[0000-0002-1590-8551]{Charlie Conroy}
\affiliation{Center for Astrophysics $|$ Harvard \& Smithsonian, 60 Garden Street, Cambridge, MA 02138, USA}

\author[0000-0001-5082-9536]{Yuan-Sen Ting (丁源森)}
\affiliation{Institute for Advanced Study, Princeton, NJ 08540, USA}
\affiliation{Department of Astrophysical Sciences, Princeton University, Princeton, NJ 08544, USA}
\affiliation{Observatories of the Carnegie Institution of Washington, 813 Santa Barbara Street, Pasadena, CA 91101, USA}
\affiliation{Research School of Astronomy and Astrophysics, Mount Stromlo Observatory, Cotter Road, Weston Creek, ACT 2611, Canberra, Australia}

\begin{abstract}
 In this paper we present a holistic view of the detection, characterization, and origin of stellar streams in the disk of a simulated Milky Way-like galaxy. The star-by-star simulation of the Galaxy evolves stars born in clusters in a realistic galactic potential that includes spiral arms, a bar, and giant molecular clouds over $5$ Gyr. We first devise a new hybrid method to detect stellar streams that combines phase space density information along with the action-angle space spanned by stars in our simulation. We find that streams' progenitor star clusters and associations are all preferentially higher-mass ($>1000$ $M_{\odot}$) and young ($< 1$ Gyr). Our stream-finding method predicts that we might be able to find anywhere from $1$ to $10$ streams with 6D \textit{Gaia} DR2 data in the solar neighborhood alone. The simulation suggests that streams are sensitive to the initial dynamical state of clusters, accumulated energy gain from encounters with giant molecular clouds (GMCs), and present-day actions. We investigate what we can learn about the Galactic potential by studying the feasiblity of rewinding stellar streams back to their origin. Even with perfect information about the non-axisymmetric components (spiral arms, bar) of the galactic potential, the stochastic GMC population makes backwards integration impossible beyond one or two disk orbital times. Streams are also sensitive to the properties of the bar, but fairly insensitive to the properties of the non-transient two-armed spiral in our simulation. Finally we predict that around $10$ to $30$ stellar streams should be detectable with \textit{Gaia}'s 10-year end-of-mission data. There are many more stellar streams waiting to be discovered in the Galactic disk, and they could hold clues about the history of the Galaxy for the past Gyr. 
\end{abstract}


\section{Introduction} 
\label{sec:intro}

The history of the Galaxy is encoded in the distribution of the kinematics and the chemistry of its stars. The deluge of astrometric \citep{brown2018gaia} and spectroscopic \citep[e.g.,][]{kollmeier2017sdss, kunder2017radial, buder2019galah, ahumada2019sixteenth} data expected in the coming years is likely to bring about new insights about how galaxies like ours are assembled over cosmic time. 

The study of stellar streams of stars in the Galactic halo in particular has shown the potential of these datasets in unraveling the early history of the Milky Way and reveal fundamental insights about the nature of dark matter \citep[for e.g.,][and references therein]{helmi2020streams}. Given their cold kinematics, small perturbation from interactions with dark matter subhaloes or the Galactic bar could potentially leave a large imprint on the dynamical signature of streams \citep[e.g.,][]{pearson2017gaps, bonaca2020high}. The kinematics and chemistry of these stars has helped us piece together the early history of the stellar halo of the Galaxy. 

The potential utility of stellar streams in the Galactic disk from disrupting open clusters or stellar associations on the other hand is largely unexplored. There are two difficulties with assessing what we can learn from stellar streams in the Galactic disk. First, perturbations from the non-axisymmetric components of the Galaxy (spiral arms and bar) and from GMCs are far more likely on disk-like orbits than halo-like orbits. Secondly, there is no clear theoretical motivation for how long stellar streams in the disk will be detectable in phase space due to the much shorter dynamical time and the complex dynamical signature of stellar feedback and gas expulsion \citep[e.g.,][]{dinnbier2020tidal}. While moving groups were originally thought to be disrupted clusters \citep[e.g.,][]{1965gast.book..111E}, most moving groups are now thought to be overdensities due to the resonances in the Galactic potential \citep[e.g.,][]{1998AJ....115.2384D}. Consequently, there is no real theoretical expectation for how many stellar streams we can hope to detect with current and future data.

The first chemically homogeneous stellar stream in the disk was recently discovered in \citep{meingast2019extended}. The stream, known as Pisces-Eridanus, is $\sim$ $120$ Myr old \citep{curtis2019tess} and extends hundreds of parsecs in the local solar neighborhood and likely contains hundreds of stars \citep{ratzenbock2020extended}. Moreover, recent work has also detected the tidal tails of nearby open clusters \citep[e.g.,][]{roser2019hyades, meingast2019extended3, roser2019praesepe, oh2020kinematic, meingast2021extended, jerabkova2021800} and a stream originated from a past satellite interaction \citep[e.g.,][]{laporte2020chemodynamical}. These detections further beg the question of what we can expect to learn about the nature of star formation, the chemical evolution of stars, and the large-scale mass distribution of the Galaxy from stellar streams in the disk. 

In \citet[][hereafter K19a]{k19a}, we presented a star-by-star dynamical model of the Galactic disk that takes into account the clustered nature of star formation and the complexity of the Galactic potential. Each resolution element in the simulation represents an individual star. The simulation self-consistently evolves 4 billion stars over the last 5 Gyr in a realistic potential that includes an axisymmetric component, spiral arms, a bar, and GMCs. All stars are born in clusters with an analytical model for cluster birth and dissolution \citep{lada2003embedded}. The simulation provides the ideal sandbox to investigate how we can find, characterize, and learn about the Galaxy from stellar streams.

The paper is structured as follows. In Section \ref{sec:ds} we discuss the simulation from \citet{k19a}, how we create mock \textit{Gaia} catalogs, and the method we use to detect stellar streams in the simulated galaxy. Section \ref{sec:survival} discusses why some stellar streams survive and remain visible in phase space and some do not. In Section \ref{sec:learn} we examine what we can learn about the Galactic potential from stellar streams. We discuss the future of stellar streams in the disk in Section \ref{sec:future} and conclude in Section \ref{sec:conc}.  

\section{Simulations}
\label{sec:ds}

\subsection{Simulations}
\label{sec:sims}

In \citet{k19a} we presented three simulations of the Galactic disk that explore different paramterizations for clustered star formation and the potential of the Galaxy. In this work we will focus on the ``fiduical" simulation presented in K19a. The fiducial simulation includes a comprehensive treatment of the clustered nature of star formation and includes a realistic potential with a bar, spiral arms, and GMCs. The fiducial simulation self-consistently evolves 4 billion stars over the last 5 Gyr in a time-varying potential that includes an axisymmetric component, a bar, spiral arms, and live giant molecular clouds (GMCs). All stars' positions and velocities are initialized with an  observationally motivated recipe for clustered star formation.

Running direct \textit{N}-body simulations for millions of star clusters comprising billions of stars is currently computationally infeasible. In K19a we presented a subgrid model for how to mimic star cluster initialization, evolution, and disruption based on analytic theory and  hydrodynamical \& \textit{N}-body simulations \citep{lada2003embedded, fujii2016formation}. Our model is calibrated to broadly agree with the cluster formation efficiency of a Milky Way-like galaxy and the evolution of the mass-radius relation for young clusters. We assume that stars older than $5$ Gyr are fully phase-mixed in the galaxy, and thereby include a smooth background to match observations of the local stellar surface density \citep{rix2013milky} and recent results on radial migration \citep{frankel2018measuring}. 

To enable a fair comparison with observational data we create mock catalogs of the fiducial simulation that mimic the spatial volume and errors of current and future \textit{Gaia} data releases. The solar neighborhood sphere is centered at $(-8.2, 0.0, 0.025)$ kpc \citep{bland2016galaxy}. The \textit{Gaia} DR2 mock catalog sphere has a radius of $0.5$ kpc and the \textit{Gaia} end-of-mission 10-year (EOM) mock catalog has a radius of $1.5$ kpc. These values were roughly chosen to indicate the spatial extent to which \textit{Gaia} parallaxes will be of high fidelity. We use the \texttt{MIST} stellar evolutionary tracks \citep{choi2016mesa} and the C3K stellar library (Conroy et al., unpublished) to derive photometry for the simulated stars using a Kroupa IMF \citep{kroupa2001variation}. We also calculate $G_\mathrm{RVS}$ using $(G-G_{RP})$ color from Equations 2 and 3 presented in \citet{brown2018gaia}, and apply the $G_\mathrm{RVS}$ selection of $G_\mathrm{RVS} < 12$ for the DR2 mock catalog and $G_\mathrm{RVS} < 14$ for the EOM mock catalog. 

\begin{figure*}
 \includegraphics[width=168mm]{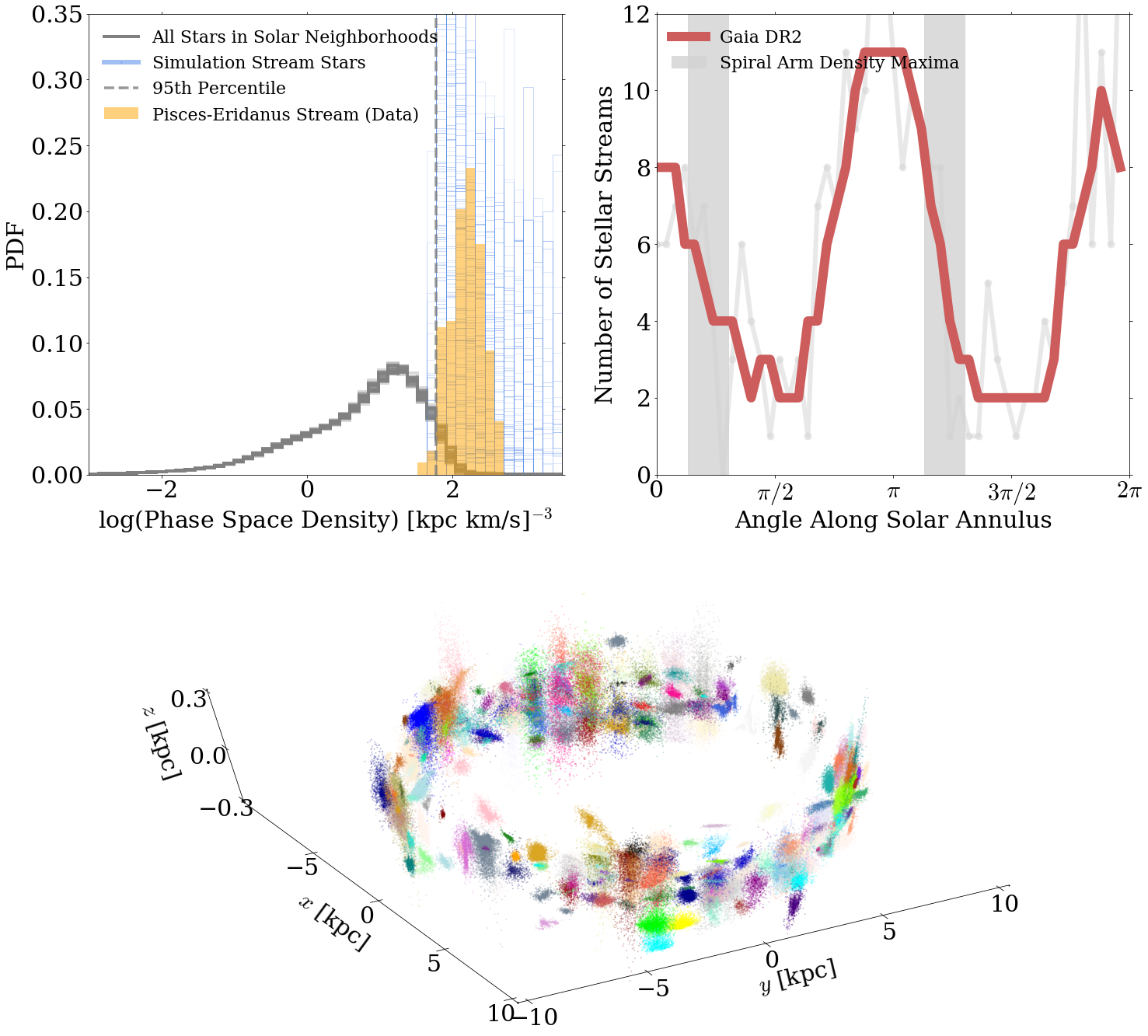}
 \caption{Stellar streams in the simulated galaxy's disk. \textit{Top-left:} The different grey histograms show the phase space density distribution for different mock catalogs in the solar annulus. Blue histograms show the streams that remain overdense in phase space. Orange histogram shows the phase space density distribution of the Pisces-Eridanus stream \citep{meingast2019extended} detected in \textit{Gaia} DR2. The grey dashed line shows the $95$th percentile in the phase space density distribution. \textit{Top-right:} Number of detected streams in \textit{Gaia} DR2-like mock catalogs in the simulation. The grey dashed line in the back shows the raw number per mock catalog and the solid red line shows the rolling average with a window size of 4. The number of streams clearly follows a sinusoidal pattern and seems to be correlated with the spiral arm density maxima. \textit{Bottom:} 3D distribution of the stellar streams in galactocentric coordinates. Each star is given a unique color based on its progenitor star cluster or stellar association.}
 \label{fig:streams}
\end{figure*}

Forecasting the number of detectable streams for \textit{Gaia} DR2 and future data releases also requires a thorough treatment of the astrometric and spectroscopic error propagation in the simulation. The dependence of parallax, proper motion and radial velocity errors is a complex function of several parameters. We follow the procedure set out in \citet{k19b, k20} to build the error model for the simulations. Briefly, we fit a Gaussian mixture model (GMM) with 20 components to the combined ($G, G_{\rm{BP}}-G_{\rm{RP}}, \sigma_{\varpi}, \sigma_{\rm{\mu_{\alpha^{*}}}}, \sigma_{\rm{\mu_{\delta}}}$) and ($G, G_{\rm{BP}}-G_{\rm{RP}}, \sigma_{\rm{RV}}$) spaces respectively for both DR2 and EOM errors, where $\sigma_{\varpi}, \sigma_{\rm{\mu_{\alpha^{*}}}}, \sigma_{\rm{\mu_{\delta}}}, \sigma_{\rm{RV}}$ are the uncertainties in the parallax, proper motions and the radial velocities. Using the $G, G_{\rm{BP}}, G_{\rm{RP}}$ values derived for each star using the procedure described above, we sample from the conditional distributions for the respective astrometric and spectroscopic errors given $G$ and $G_{\rm{BP}}-G_{\rm{RP}}$. The scanning law also has an impact on the selection of stars in the solar neighborhood and their respective errors. With recent progress on modelling the scanning law and computing the true RVS selection function \citep{boubert2020completeness}, we plan to include a detailed selection function in future work (as opposed to a simple magnitude cut) for both the error model and the selection of stars.

\subsection{Finding \& Characterizing Streams in the Simulation}
\label{sec:finding}

Only one conatal stellar stream has been definitively detected in the Galactic disk solely using dynamical properties of stars \citep{meingast2019extended}. There is also some related recent work that suggests the existence of other such stellar streams but these results have not been validating using chemical information or astroseismology \citep{kounkel2019untangling}. The observational detection of the Pisces-Eridanus stream was performed using a combination of wavelet analysis and DBSCAN. The wavelet decomposition constructs a series of components at different velocity scales -- the authors focus on velocity amplitudes of $\sim 1.5$ km s$^{-1}$ to identify the overdensity and then perform a clustering algorithm, DBSCAN, to select stream stars. Both analysis steps here require hand-tuning and subsequent analysis to identify a ``clean" sample of stream stars. 

We adopt a different approach in this work to minimize hand-tuning and utilize the information content of action-angle coordinates \citep[e.g.,][]{trick2019galactic}. We first use the coarse grained phase space density ($f$) to quantify the relative overdensity of a given star in phase space in our simulation and then perform clustering in action-angle space for stars overdense in phase space. $f$ is the density in a finite six-dimensional volume defined by $d^3x d^3v$, centered on the phase space position of a given star at $(x,v)$. Phase space densities have been extensively used to study substructure in galaxy haloes \citep[e.g.,][]{helmi2002phase, hoffman2007evolution}, but their use in the disk has been limited. The results presented in this paper utilize EnBID \citep{sharma2006multidimensional}, which builds upon \citet{ascasibar2005numerical}, to numerically calculate the phase space densities using a binary tree and an entropy-based splitting criterion.

Once the phase space density for each star has been calculated, we select overdense stars above a certain threshold. Since the phase space density alone cannot partition overdense stars into constituent streams, we calcualate the actions and angles for these overdense stars using Agama \citep{vasiliev2019agama} and perform clustering in action-angle space using HDBSCAN.  

The detailed procedure to identify streams for the purposes of this work is listed below. 

\begin{enumerate}
\item Calculate the phase space density ($f$) of all stars in a given solar neighborhood mock catalog using EnBID \citep{sharma2006multidimensional}
\item Select all stars in the given solar neighborhood that have $f$ $\geq 95$th percentile
\item Calculate actions and angles for these stars overdense in phase space using Agama \citep{vasiliev2019agama}
\item Run HDBSCAN \citep{mcinnes2017hdbscan} on the calculated actions and angles 
\item For each detected clump by HDBSCAN, if the purity (i.e. the number of stars from the dominant star cluster in the clump) is $>50\%$ and the number of stars $>50$, we consider this to be a detected stellar stream
\end{enumerate}

About $\sim 15 - 30$ \% of all clumps that HDBSCAN finds at step 4 are real detections of conatal streams. Consequently, the key caveat in our detection method is that the purity cut requires independent information to confirm conatality. In the simulations, the birth cluster of each star is known a priori and makes the calculation of ``purity" (i.e., how many stars in a given clump were born from the same cluster) of each HDBSCAN-detected clump trivial. However, performing the same calculation observationally would require some other way to prove conatality. As shown in \citet{hawkins2020chemical}, spectroscopic follow-up for candidate streams can strongly indicate a common origin by demonstrating chemical homogeneity \citep[e.g.,][]{bovy2016chemical}. Furthermore, future spectroscopic surveys will also likely provide chemical information for tens of millions of stars contiguously in the Galactic disk \citep[e.g.,][]{kollmeier2017sdss, kunder2017radial, buder2019galah, ahumada2019sixteenth}. The stream detection method above will be tested out on future \textit{Gaia} and spectroscopic data in future work (Kamdar et al. in prep). 

The top-left panel of Figure \ref{fig:streams} offers a visual demonstration of how we detect streams in the simulation. The different black lines show the phase space density  distribution of all stars in the different solar neighborhoods in the fiducial simulation. The dashed grey line shows the $95$th percentile of the phase space density distribution of stars in one of the solar neighborhoods. The different blue histograms show the phase space density distribution for stars in our simulation that we detect as streams. The Pisces-Eridanus stream's phase space density distribution is shown in orange. The similarity between the simulation streams and the observationally validated Pisces-Eridanus stream further solidify our method of detecting streams in the simulation. 

The grey dash-dotted line in the top-right panel of Figure \ref{fig:streams} shows the number of detected streams in different solar neighborhoods as a function of the solar neighborhood's azimuthal angle along in the simulated galaxy. The red line is a smoothed running mean with a window size of 5 points for the number of streams. The shaded grey regions show the maxima in the density of the two-armed spiral pattern present in our Galactic potential. The number of streams clearly follow a sinusodial pattern as a function of the azimuthal angle of the solar neighborhood in the galaxy. 

The prescription used to initialize stars in \citet{k19a} used the maxima in spiral arms' density to ensure that the bulk of star formation occured within $\pm 30$ degrees of the spiral arms. Since most streams are fairly young, the correlation with the spiral arms' position is reasonable since the stars' memory of their birthplace is not fully lost. The streams are likely offset due to the difference in the pattern speed for the spiral arms ($25$ km s$^{-1}$ kpc$^{-1}$, which translates to an orbital period of about $\sim 250$ Myr) and the circular velocity at the solar location ($\sim 240$ km s$^{-1}$, which translates to an orbital period of about $\sim 205$ Myr). 

The bottom panel of Figure \ref{fig:streams} shows a 3D view of the fiducial simulation with different stellar streams in unique colors. The different azimuthal angles along the solar annulus are labeled near the top of the panel. The sinusodial pattern in the number of streams is clear in this panel as well. Streams in the simulation shown in the Figure bear a similarity to the bead-like structure observed in \citet{kounkel2019untangling}, though in a larger spatial volume. 

\section{What Makes A Stream?}
\label{sec:survival}

Since there has only been one detection of an unbound stellar stream in the disk so far, it is worth asking why some streams stay overdense (hence detectable) in phase space whereas most do not. This question requires a suitable control sample for comparison. To that end, we use importance sampling to select stars born in clusters that follow the same mass and age function as the streams but are not detected as streams using the procedure described in Section \ref{sec:finding}. The stars in these control clusters are of similar masses and ages by design but are essentially phase mixed and undetectable in phase space alone. 

The current overdensity of stars in phase space that we detect as streams could feasibly have four different explanations:
\begin{itemize}
    \item initial dynamical state of the star cluster 
    \item initial orbital properties
    \item current orbital properties that make it easier to disentangle from background stars
    \item the amount of scattering or energy injection along a star cluster's orbit
\end{itemize}

In this section we analyze these different mechanisms to understand why some stars remain overdense in phase space up to a Gyr. 

\subsection{The Impact of Birth Conditions}
\label{sec:sf}

\begin{figure*}
 \includegraphics[width=168mm]{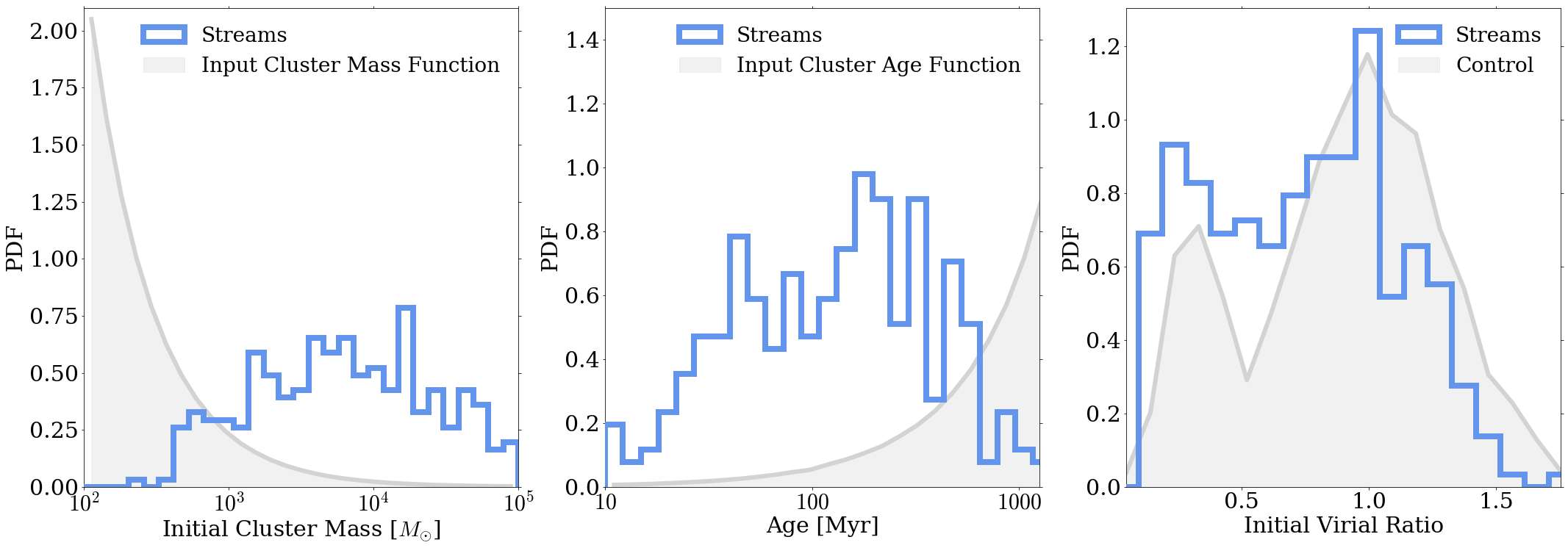}
 \caption{Characterizing the massses, ages, and initial dynamical state of stellar streams detected in the simulation. Streams preferentially originate in higher-mass star clusters / associations and are mostly younger than $1$ Gyr. \textit{Left panel:} Input cluster mass function (grey) and stream progenitor masses (blue). \textit{Middle panel:} Input cluster age function (grey) and ages of streams (blue). \textit{Right panel: } Initial dynamical state for stream stars (blue) and control stars (grey). The virial ratio with which star clusters are initialized controls their initial boundedness. The bimodality in both distributions is from the subgrid model from \citet{k19a} where some the majority of star clusters are born overvirialized to mimic gas expulsion from stellar feedback. The distributions are fairly similar with stream progenitors being slightly more likely to be dynamically bound at birth.}
 \label{fig:stream_props}
\end{figure*}

Figure \ref{fig:stream_props} characterizes the mass, age, and initial dynamical state of the progenitor star clusters and associations for detected stellar streams in the simulation. The left panel shows the streams' birth cluster masses and the input cluster mass function. As a reminder we impose a minimum of $50$ stars per detected stream. The panel clearly shows that the star clusters that produce streams have a preferentially far larger mass compared to the overall simulation. The right panel shows the streams' birth cluster ages and the input cluster age function (i.e., star formation history). Streams are largely all younger than $1$ Gyr. This is similar to the result we found in \citep{k19a}, where we showed that phase space of the Galactic disk only remains clumpy for about a Gyr. 

In K19a we implemented a subgrid model for gas expulsion and cluster disruption due to stellar feedback by intentionally overvirializing a majority of birth clusters. The right panel of Figure \ref{fig:stream_props} compares the initial virialization of streams' progenitor clusters (blue) and control clusters (grey). The bimodality in both distributions is due to the prescription for gas expulsion, where some star clusters are initialized as bound but the majoirity as unbound. This panel shows that the initial dynamical state of streams' progenitors is slightly more bound compared to the control stars' progenitor clusters. This initial boundedness likely translates into a higher probability of detectability at present day. However, the large overlap in these distributions suggest that the simulation's prescription for gas expulsion alone could not explain why some star clusters produce detectable streams and some do not.

\subsection{Initial Dynamical Properties}
\label{sec:properties}

Figure \ref{fig:birth} shows different dynamical properties of the progenitor star clusters and associations that eventually produce stellar streams. The stars belonging to the control sample are colored grey and stars belonging to streams are colored blue. The top two panels show the birth galactocentric radius ($R$; right panel) and the height off the midplane ($|Z|$; left panel). The bottom two panels show the birth vertical action, $J_z$, and birth radial action, $J_r$. 

The birth $|Z|$ and $R$ distributions look fairly similar for the stream stars and the control stars. The chosen pattern speeds for the spiral arms and bar in the non-axisymmetric components of the simulation lead to a resonance overlap at $\sim 8.9$ kpc \citep{k19a}. The bimodality in the streams' birth $R$ distributions could perhaps indicate that star clusters born near resonances could be more likely to create stellar streams in the disk. 

We do not observe stream or control stars born from much further away than $1-1.5$ kpc away from the solar location. This is largely driven by the small spatial volume of the DR2 mock catalog ($0.5$ kpc sphere). Since the control and stream samples both contain young stars by design, the amount of radial migration they could have experienced in the simulation is fairly low -- empirical estimates for the radial migration of similar stars from \citet{frankel2018measuring} would indicate a dispersion of about $\sim 1.3$ kpc in a Gyr. The birth $J_z$ and $J_r$ distributions also look largely identical. There is a slight hint of the radial action having a longer tail for stream stars, but the signal is not strong due to the small sample size. 

\subsection{Current Properties} 
\label{sec:current}

\begin{figure}[t!]
 \includegraphics[width=0.45\textwidth]{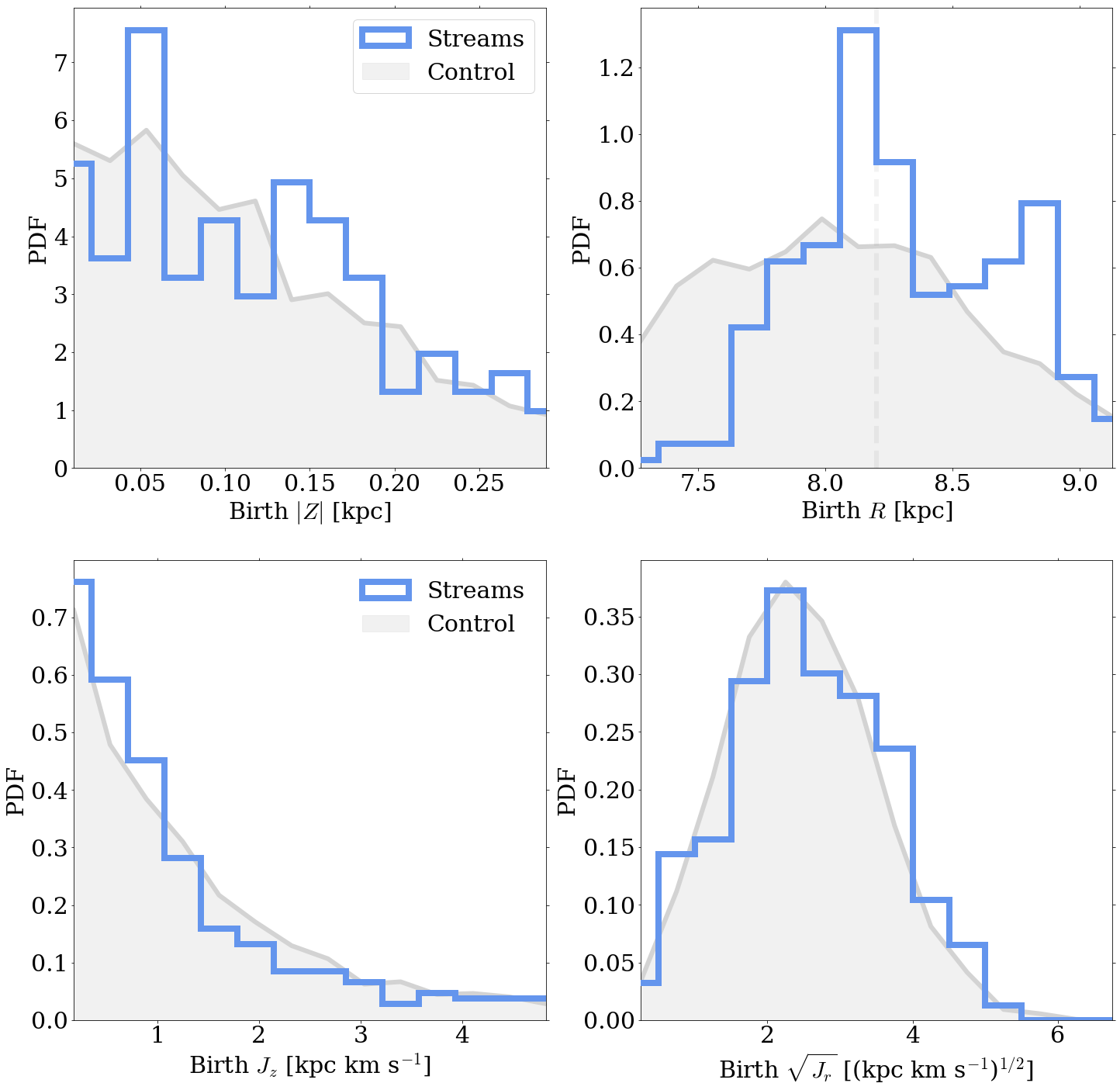}
 \caption{Birth properties of stream and control stars' progenitor clusters. Stream stars in blue and control stars in grey. \textit{Top-left:} Birth height off the midplane ($|Z|$) distribution for the two different samples of stars. \textit{Top-right:} Birth galactocentric radius ($R$) for stream and control stars. The distributions are similar. \textit{Bottom-left:} Vertical action distribution ($J_z$). \textit{Bottom-right:} Radial action ($\sqrt{J_r}$) distributions.}
 \label{fig:birth}
\end{figure}

\begin{figure}[!t]
 \includegraphics[width=0.45\textwidth]{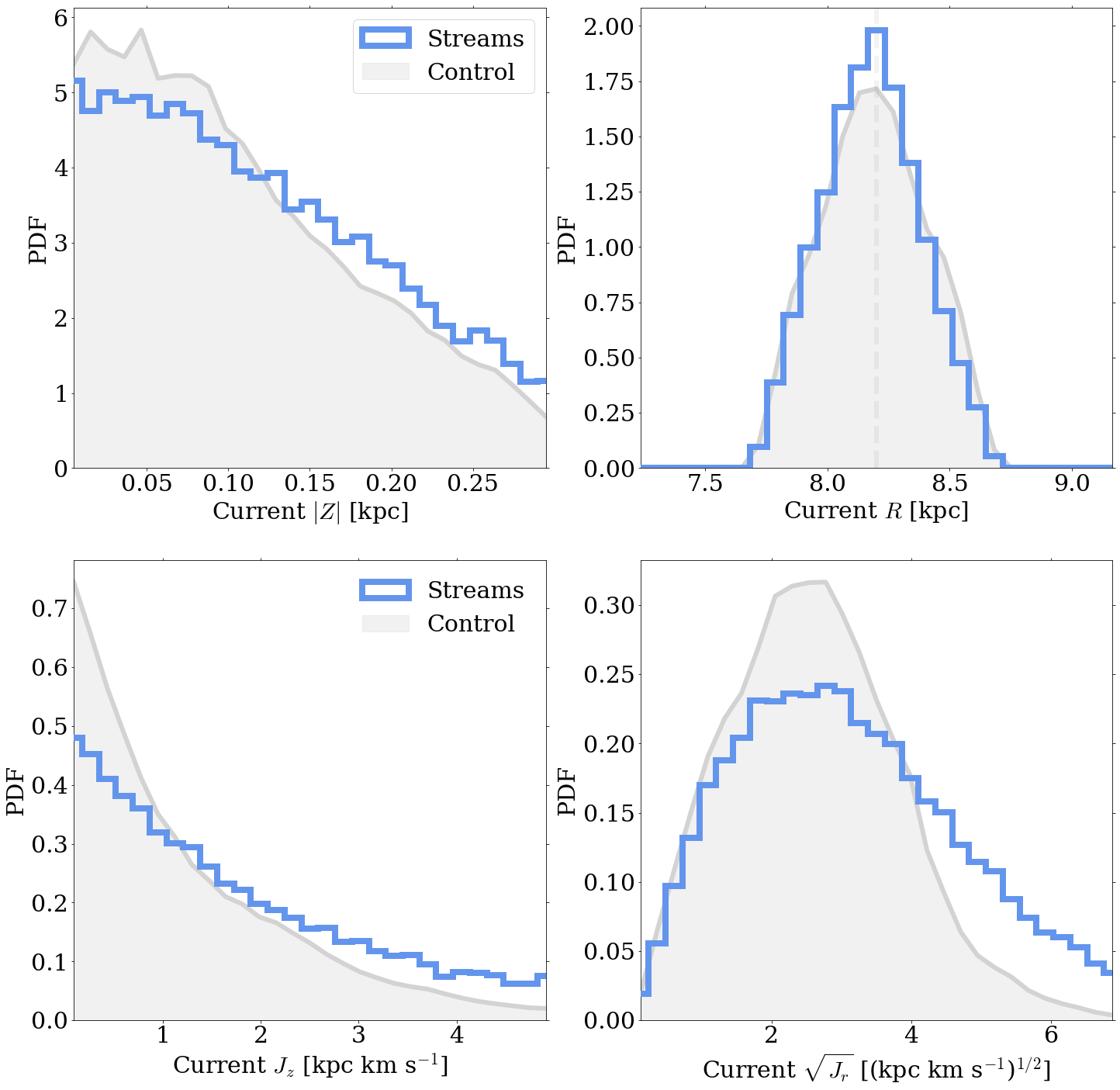}
 \caption{Same as Figure \ref{fig:birth} but for current properties of stream and control stars. }
 \label{fig:current}
\end{figure}

Figure \ref{fig:current} shows different dynamical properties at the present time for stream stars and control stars. It is structured similarly to Figure \ref{fig:birth} -- the top two panels show $|Z|$ and $R$, and the bottom two panels show $J_z$ and $J_r$. The blue histograms correspond to stream stars' properties and the grey histograms correspond to the control sample's properties. 

The same broad patterns are imprinted in the current dynamical properties of stream stars as their progenitors that we saw in Figure \ref{fig:birth}. The current $|Z|$ distribution for stream stars is slightly broader than the control and indicates that stream stars are more likely to be further away from the midplane. The histogram for the current galactocentric radii of stream stars seems to be tigher near the sun's location compared to the control sample. The narrower distribution here could be reflecting that it is easier to pick out overdense stars closer to us because the astrometric and spectroscopic errors for nearby stars is smaller. 

Both the radial action and the vertical action for stream stars have a noticeably longer tail than the control sample. The larger actions indicate that they are on slightly more unique orbits compared to the field population. This seems reasonable since our detection method prioritizes locally overdense stars in phase space -- one would expect to pick out stars on unique orbits due to the lower background phase space density. The distributions, while notably different, are still fairly similar. Therefore, the present-day dynamical properties of stellar streams seem to have a small but notable difference from the control sample. However, they do not seem to solely determine why some stars remain overdense in phase space and some do not.

\subsection{The Impact of GMCs}
\label{sec:gmcs}

\begin{figure}
 \includegraphics[width=84mm]{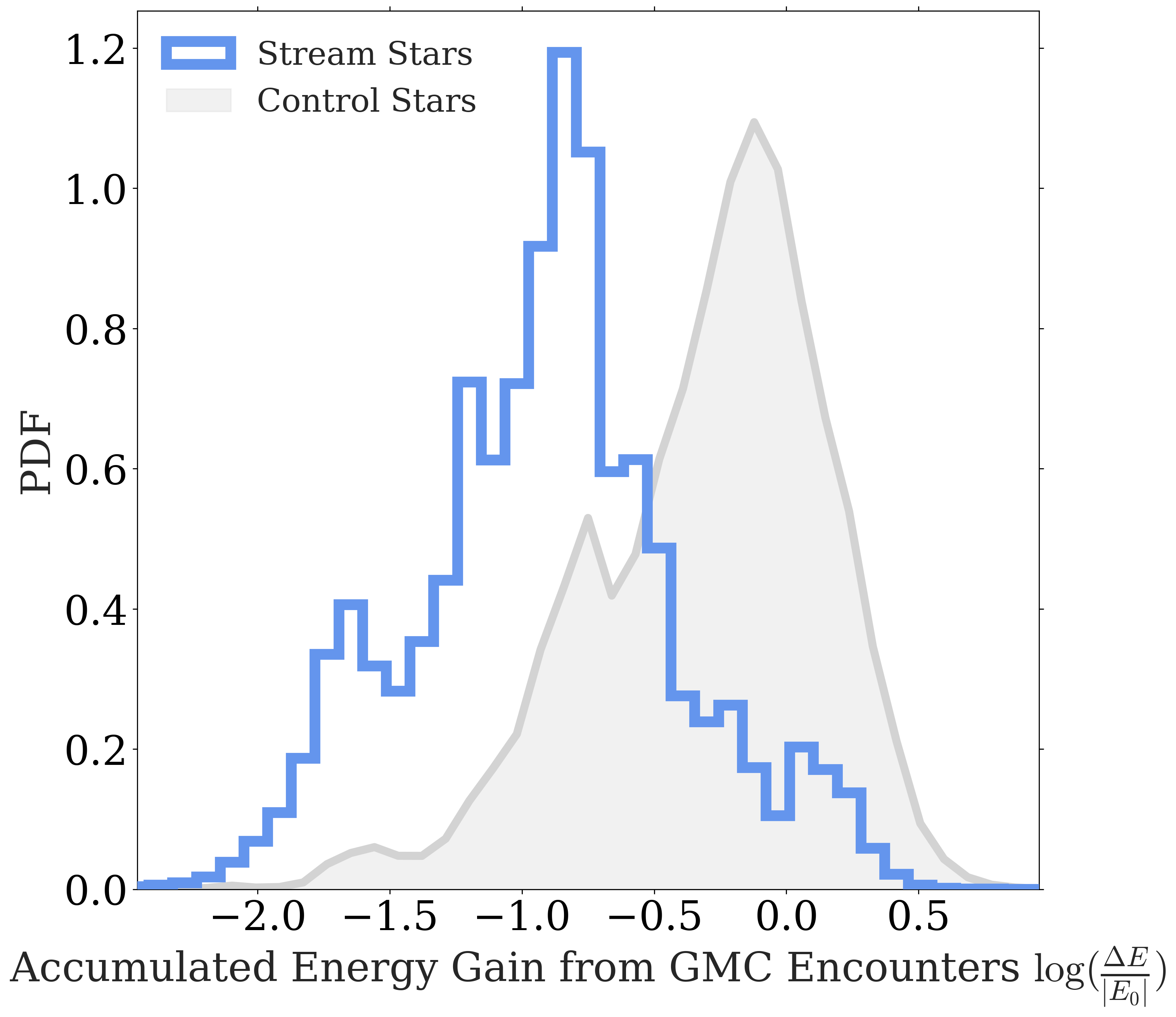}
 \caption{Accumulated energy gain ($\log \frac{|\Delta E|}{E_0}$) from GMC encounters along stars' orbits. Stream stars are shown in blue and control stars are shown in grey. Stellar streams gain almost an order-of-magnitude less energy compared to control stars.}
 \label{fig:gmcs}
\end{figure}

Interactions between star clusters and GMCs are known to be destructive \citep[e.g.,][]{1958ApJ...127...17S, gieles2006star} and have been used in the past to explain the dearth of old open clusters in the solar neighborhood. Relatedly, recent work \citep[e.g.,][]{ting2019vertical} has also shown the efficiency of GMC scattering explains the vertical heating history of stars in the thin disk of the Milky Way. \citet[][hereafter G06]{gieles2006star} show that the approximate disruption time for an initial star cluster with mass $1000$ M$_{\odot}$ in the solar neighborhood with a uniformly distributed GMC background is about $\sim 100-200$ Myr. The simulations in G06 suggest that young clusters are destroyed by just a few chance GMC encounters. Therefore, individual birth clusters might have significantly different lifetimes due to the inherent stochasticity of GMC birth, evolution, and dispersal. 

Figure \ref{fig:gmcs} shows the difference between the total energy gain ($\log \frac{|\Delta E|}{E_0}$) from all GMC encounters for stream stars (in blue) and for control stars (in grey) from their birth to the present day. The two arrows near the top of the panel mark the respective peaks of the two distributions. The difference between the peaks of the accumulated energy gain distributions for these two samples is almost an order of magnitude. For reference, a single encounter with a GMC of mass $10^5$ M$_{\odot}$ and radius $\sim 15$ pc, at an impact parameter of $\sim 10$ pc and a relative velocity difference of $\sim 10$ km s$^{-1}$ would roughly correspond to $\frac{|\Delta E|}{E_0} \sim 1$ and be sufficient to disrupt the stream. 

The overdensity of stream stars, then, is driven by three factors -- they were initially born in a dense environment with a higher likelihood of initial boundedness, these stars have not yet encountered any large scatterers on their orbit, and their present-day dynamical properties seem easier to disentangle from the background. The largest difference between stream stars and control stars' properties is observed in the difference in accumulated energy from GMC encounters, which suggests that GMCs might be the primary determinant in whether or not stars remain detectable in phase space as stellar streams.

\section{What Can We Learn From Streams?}
\label{sec:learn}

\subsection{Galactic Potential}
\label{sec:potential}

\begin{figure}[t!]
 \includegraphics[width=84mm]{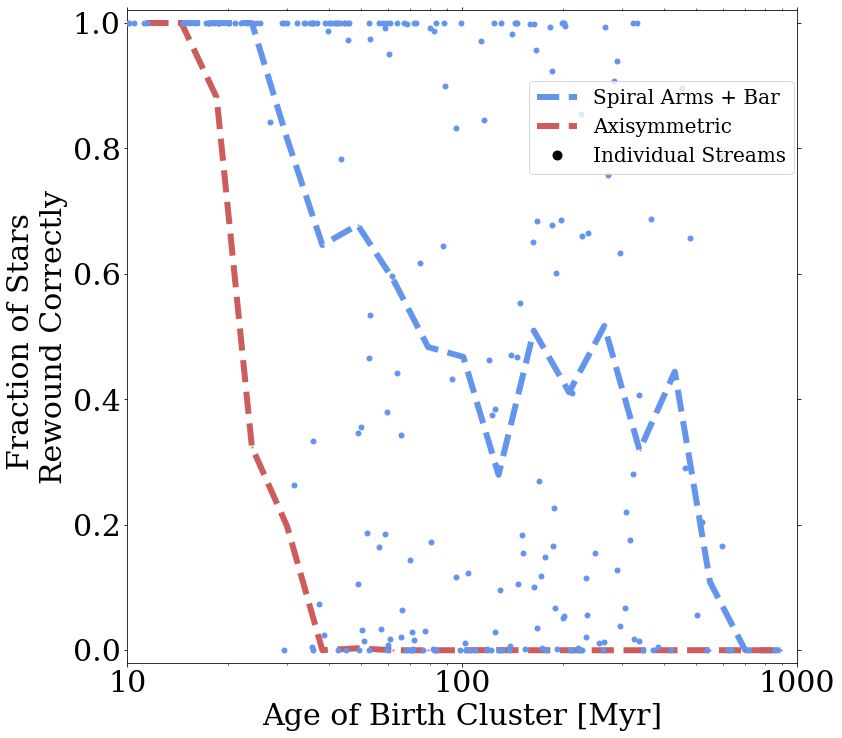}
 \caption{Backward integration of stellar streams with perfect information about different components of the potential but no information about the history of GMCs in the simulation. The blue line shows the fraction of stars rewound correctly (i.e., within $5$ pc of their birth location) when the spiral arms and the bar are included in the backwards integration. The red line shows the fraction of stars rewound correctly with only the axisymmetric component of the potential.}
 \label{fig:reverse}
\end{figure}

\begin{figure*}
 \includegraphics[width=168mm]{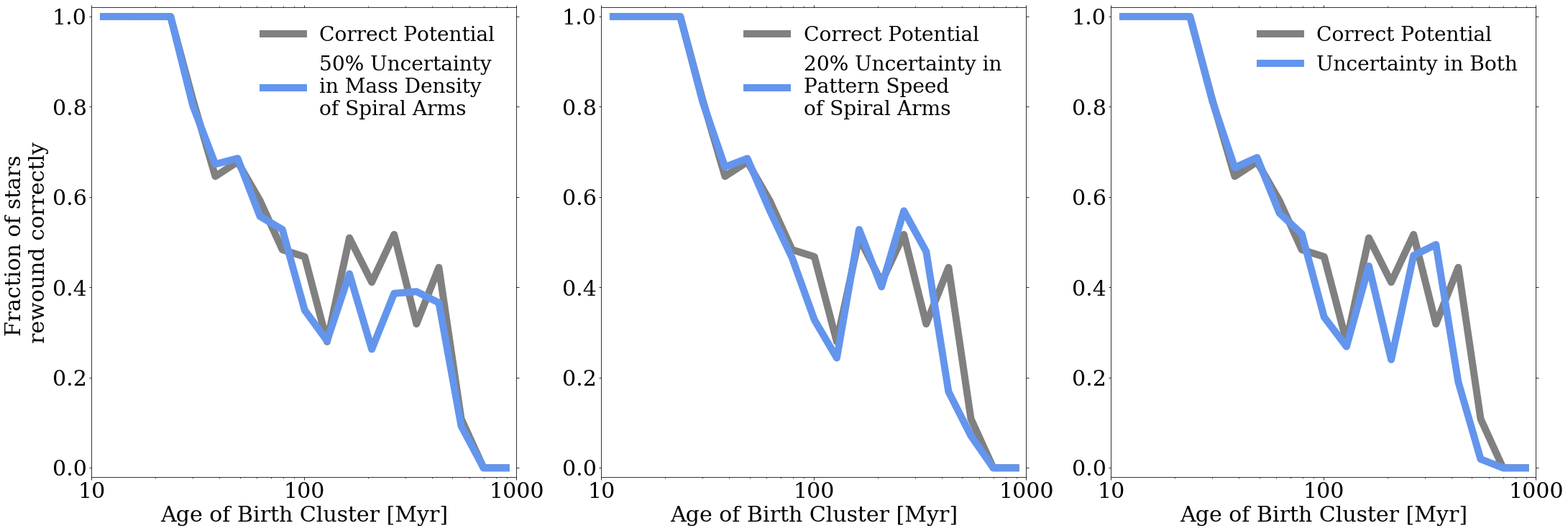}
 \caption{Impact of uncertainties in the spiral arm parameters on the fraction of stars rewound correctly. The different panels vary the mass density, pattern speed, or both (blue lines) and compare with the backwards integration with perfect information from Figure \ref{fig:reverse}. The relative insensitivity of the backwards integration to the spiral arms' parameters suggest that it would be hard to constrain the history of spiral structure using stellar streams. }
 \label{fig:spiral}
\end{figure*}

\begin{figure*}
 \includegraphics[width=168mm]{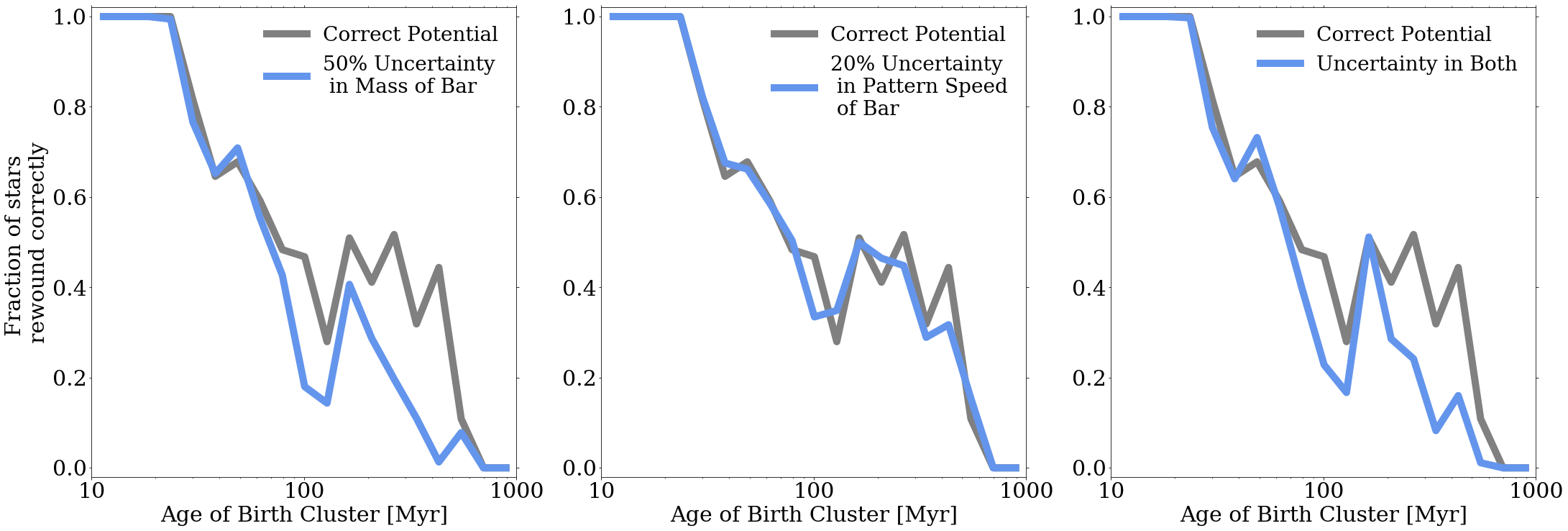}
 \caption{Same as Figure \ref{fig:spiral} but with uncertainties in the parameters of the bar. The fraction of stars rewound correctly does seem to be adversely affected with uncertain values for the bar parameters. Stellar streams could potentially be used to constrain the bar's parameters. }
 \label{fig:bar}
\end{figure*}

While the discovery and characterization of conatal stellar streams in the disk would have an immediate impact on the study of gyrochronology \citep[e.g.,][]{curtis2019tess} and the chemistry of stars \citep[e.g.,][]{hawkins2020chemical}, the question of how much we can learn about the dynamical state of the disk is unclear. The mass distribution in different components of the Galaxy is of great importance to studies of the composition, history, and current dynamical properties of the Galaxy \citep[e.g.,][]{bland2016galaxy}. Stellar streams in the halo have been extensively used to constrain the potential and the density distribution of the Galactic halo \citep[e.g.,][]{malhan2019constraining, hattori2020action, reino2020galactic}. Recent work has shown that the tidal tails of Hyades are likely sensitive to the exact prescription for the Galactic potential and the presence of GMCs \citep{jerabkova2021800}.

One way to probe the potential is to assess the accuracy of rewinding the orbits of stars backwards in different potentials \citep[e.g.,][]{quillen2018spiral, crundall2019chronostar}. The destructive impact of GMCs foreshadows the difficulty in rewinding orbits back to their birthplace. Moreover, recent work \citep[e.g.,][]{beane2019implications} has also shown that local density variations in the midplane could further complicate orbit integration. The number of streams that will eventually be detected in our direct vicinity could hold intriguing clues about the history of GMC formation and evolution in the past Gyr. 

Figure \ref{fig:reverse} shows the results of backwards integration of stream stars with two different potentials. Here we define correctly as being rewound to within $5$ pc of the given star's birthplace. The blue line shows the fraction of stars rewound correctly when we provide perfect information about spiral arms and the bar, but no information about the history of the GMCs. The various blue dots show individual streams' correctly rewound fraction. The red line shows the fraction of stars rewound when only the axisymmetric portion of the potential (with the non-axisymmetric masses redistributed) is used in the backwards integration. 

The results are pessimistic even with perfect information about the input potential. Less than half the stars can be rewound correctly after a dynamical time ($\sim 220$ Myr) in the potential with spiral arms and bar. The distribution of individual streams' correctly rewound fraction seems to be fairly bimodal; some streams are very easily rewound up to a few dynamical times and some are poorly rewound only after $10-20$ Myr. The vast majority stars cannot be rewound correctly after only few tens of millions of years in the axisymmetric case. 

The bimodality in Figure \ref{fig:reverse} for different streams suggests the strong sensitivity of the backwards integration to GMC encounters. The parameters that control the encounter probability and the destructiveness of encounters between GMCs and stars are: (1) the mass function of GMCs, (2) the number density of GMCs, (3) the lifetime of GMCs, and (4) the velocity dispersion of the GMC and stellar population. Varying these parameters for the full simulation presented in \citet{k19a} is computationally expensive; however, we plan to run smaller, more controlled simulations to determine whether individual or a small collection of streams could allow us to place constraints on these various GMC properties. 

Figure \ref{fig:spiral} shows the same fraction of stars rewound correctly but with varying uncertainties in the spiral arms of the potential. The left panel shows the results for a $50$\% uncertainty in the mass density, middle panel shows $20$\% uncertainty in the pattern speed, and the right panel shows both combined. Surprisingly, these added uncertainties largely seem to make little difference on the fraction of stars rewound correctly. It is worth noting that the simulation uses a two-armed spiral pattern but recent results suggest a four-armed spiral pattern might be more appropriate. Moreover, the simulation also currently has steady-state spiral arms as opposed to transient spiral arms -- transient spiral arms would likely make backwards orbit integration even more challenging. The relative insensitivity of the backwards integration to the different spiral arm parameters would suggest that constraining the major components of the spiral arms would likely be challenging using stellar streams. 

Figure \ref{fig:bar} shows how different uncertainties in the potential of the bar impact the backwards integration of stream stars. An incorrect mass for the bar leads to a precipitous drop-off in the fraction of correctly rewound stars. Similarly, an incorrect pattern speed also seems to have a notable impact on the fraction. As one might expect, the bar seems to have a more tangible impact on the immediate history of the orbits of stars in the solar neighborhood.

\section{The Future of Streams in the Disk}
\label{sec:future}

\begin{figure}
 \includegraphics[width=84mm]{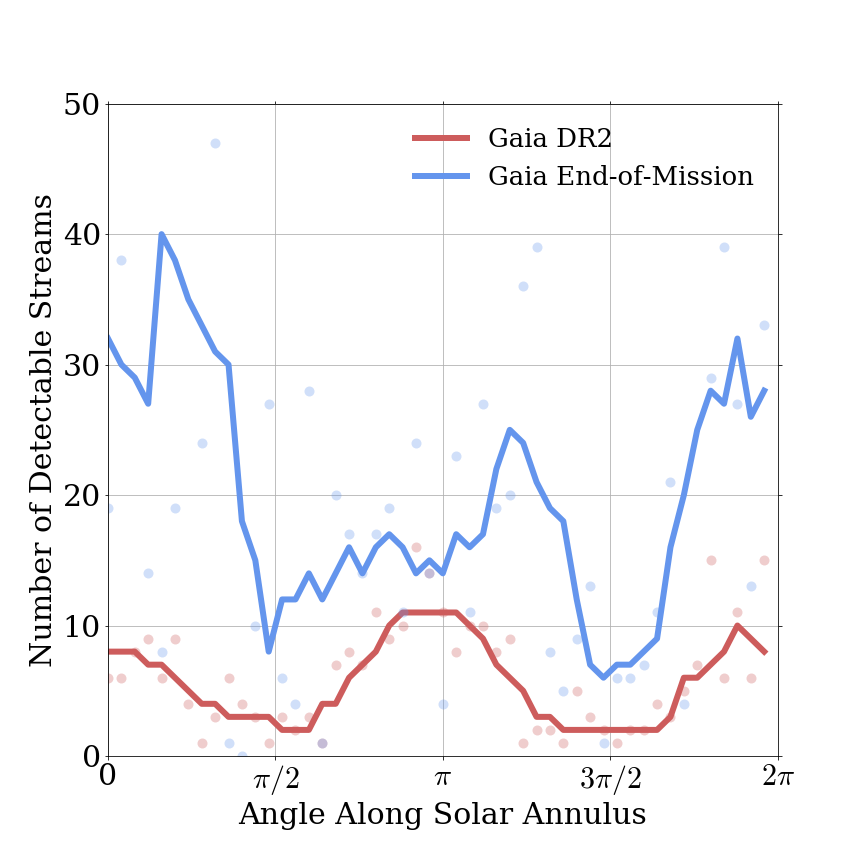}
 \caption{Number of streams as a function of the angle along the solar annulus for \textit{Gaia} DR2 and end-of-mission mock catalogs. We predict that \textit{Gaia} EOM data will likely detect a factor of $5-10$ as many streams in the Galactic disk as we see today.}
 \label{fig:nstreams}
\end{figure}

\begin{figure*}
 \includegraphics[width=168mm]{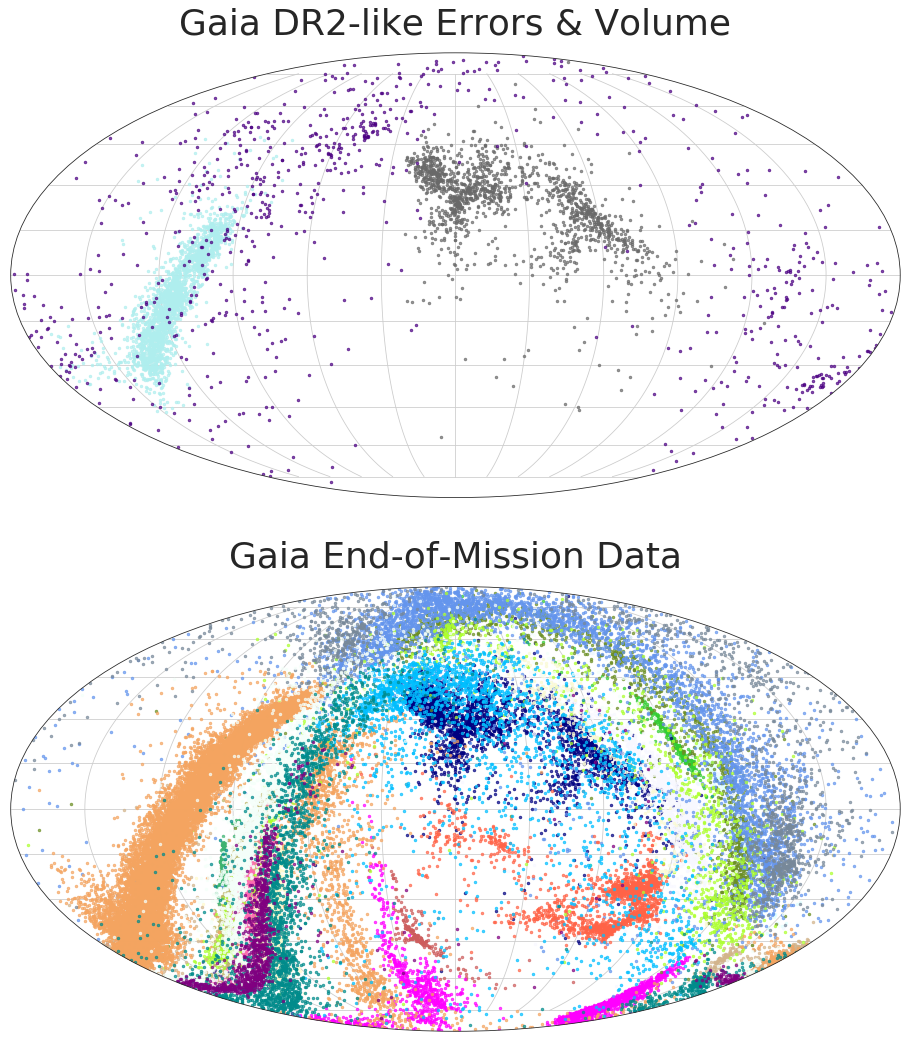}
 \caption{On-sky distribution of stream stars for a single DR2 mock catalog (top panel) compared with the same location's EOM mock catalog (bottom panel) in equatorial coordinates. The improved errors and the far larger spatial volume spanned by \textit{Gaia} EOM data will likely have an enormous impact on the future of stellar stream studies in the Galactic disk. }
 \label{fig:sky}
\end{figure*}

Future \textit{Gaia} data releases in the coming decade will deliver many more stars with radial velocities and far lower astrometric and spectroscopic errors. The \textit{Gaia} data improve with time as $t^{-0.5}$ for parallaxes, photometry, and radial velocities, and as $t^{-1.5}$ for proper motions\footnote{http://www.astro.lu.se/gaia2017/slides/Brown.pdf}. To build the mock catalogs with the error model corresponding to \textit{Gaia} end-of-mission (EOM) 10-year data, we scale the parallax and the RV errors by approximately $65\%$ and the proper motion errors by approximately $90\%$ compared to the DR2 error model. Since the parallaxes will be far more reliable further away from the solar system and we will have radial velocities at fainter magnitudes ($G_{RVS} \sim 14$ compared to $G_{RVS} \sim 12$), we also expand the volume of each mock catalog from a sphere with radius $0.5$ kpc to a sphere with radius $1.5$ kpc. 

We perform the same selection cuts presented in Section \ref{sec:finding} on the \textit{Gaia} EOM mock catalogs. Figure \ref{fig:nstreams} shows the number of detectable streams for the EOM catalogs (blue) compared to the DR2 results presented earlier in Figure \ref{fig:streams}. Regardless of the solar neighborhood's position in the simulated galaxy, we find that the number of theoretically detectable streams in the data is going to increase by a factor of $\sim 6$ on average. The same periodic pattern observed in Figure \ref{fig:streams} persists with the larger mock catalogs as well. 

We choose one of the simulated DR2 solar neighborhoods with only three stellar streams currently detectable (similar to observational data today) and compared their own-sky extent with the corresponding EOM catalog in Figure \ref{fig:sky}. As observed in \citet{meingast2019extended}, these stellar streams can occupy hundreds of degrees on the sky. The bottom panel shows the on-sky extent of the detectable streams in the EOM catalogs. 

With future \textit{Gaia} data releases, the number of stars with reliable RVs and parallaxes is likely to increase by a factor of a few. The sample could potentially be large enough to focus on the MSTO region of the CMD with reliable ages for stars and fold in that information into finding stellar streams. Moreover, contiguous spectroscopic surveys such as SDSS-V \citep{kollmeier2017sdss} provide another channel of information -- chemical abundances -- that were largely unexplored in this work. Recent work \citep[e.g.,][]{price2020strong, nelson2021distant} has shown that chemically tagging stars \citep{freeman2002new} that were born together might indeed be possible with current and future spectroscopic surveys. Combining dynamical and chemical information of stars to both find and characterize stellar streams in the disk could present the next frontier in the study of the assembly history of the Galactic disk. 

\section{Conclusion}
\label{sec:conc}

Several key physical processes including the clustered nature of star formation, non-axisymmetries of the Galactic potential, and the history of GMCs determine the fate of stars born in the Milky Way disk. The study of dynamically cold stellar streams has yielded key insights about the nature of dark matter, the potential of the Galaxy, and the physics of star clusters in the Galactic halo. However, the study of stellar streams is in its relative infancy in the Galacitc disk with only one confirmed detection and uncertainty about what we can learn about the disk from these streams. In this paper we used the simulations presented in \citet{k19a} to study the detection, characterization, and utility of stellar streams in the disk. 

The streams in our simulation are detected by analyzing overdensities in phase space. We analyzed the birth dynamical properties of streams' progenitor star clusters and the current dynamical properties of stream stars. We also show what we can learn about star formation and the galactic potential from studying these streams. Our key findings are listed below. 

\begin{itemize}
	\item We find that streams' progenitor star clusters and associations have largely the same birth $R$, $|Z|$, $J_z$, and $J_r$ as the field population. Present-day stream stars have a larger tail in both the vertical and radial actions ($J_z$ and $J_r$) compared to the control sample. Moreover, stream stars also have a slightly narrower $R$ distribution, and are slightly more likely to be further away from the midplane. 

	\item The two primary mechanism that predict whether stars remain overdense in phase space seems to be the accumulated energy gain for stars over their lifetime due to interactions with GMCs and the initial dynamical state of star clusters. Streams' progenitor clusters are more likely to be initially more bound compared to the control sample. We find almost an order of magnitude smaller accumulated energy gain from GMC encounters for stream stars compared to the field population. Consequently, only one or a few chance GMC interactions can be wholly destructive to a dynamically cold stellar stream. 

	\item Given their large impact and stochastic nature, GMCs make studying the dynamical state of the disk using stellar streams difficult. We find that less than half the stream stars in our simulation can be integrated backwards in time correctly for more than a dynanmical time ($\sim 200$ Myr) even with perfect information about the simulated galaxy's spiral arms and bar (but no information about the GMC population). We also find that our chosen parameters for steady-state, rotating spiral arms are largely insensitive to the fraction of stars rewound correctly -- consequently, constraining parameters for the spiral arms using stellar streams will likely be frought with challenges. The simulated galaxy's bar mass seems to have an outsized impact on the fraction of stars rewound correctly, which implies that we may be able to constrain the parameters of the bar using stellar streams in the future. 

	\item Future releases of \textit{Gaia} data will lead to significant improvements in the astrometric and spectroscopic errors and will provid reliable data for a much larger volume in the Galaxy. We predict that the number of stellar streams detected dynamically will likely increase by a factor of $\sim 5-10$. 

\end{itemize}

The future for Galactic archaeology and the study of long disrupted stellar streams in the disk is bright. We expect significant insights into the nature of star formation and the larger mass distribution in the Galaxy as more streams are discovered.  

\acknowledgements

We thank members of the Conroy group at Harvard for useful discussions and helpful comments. HMK thanks Harita Koya for her continued support. HMK  acknowledges  support  from  the  DOE  CSGF  under  grant  number DE-FG02-97ER25308. CC acknowledges support from the Packard Foundation. YST is supported  by the NASA Hubble Fellowship grant HST-HF2-51425.001 awarded by the Space Telescope Science Institute. The computations in this paper were run on the Odyssey cluster supported by the FAS Division of Science, Research Computing Group at Harvard University. 

This work has made use of data from the European Space Agency mission \textit{Gaia} (\url{https://www.cosmos.esa.int/gaia}), processed by the Gaia Data Processing and Analysis Consortium (DPAC, \url{https://www.cosmos.esa.int/web/gaia/dpac/consortium}). Funding for the DPAC has been provided by national institutions, in particular the institutions participating in the Gaia Multilateral Agreement. The Sloan Digital Sky Survey IV is funded by the Alfred P.Sloan Foundation, the U.S. Department of Energy Office of Science, and the Participating Institutions and acknowledges support and resources from the Center for High-Performance Computing at the University of Utah. 

{\it Software:} \texttt{IPython} \citep{perez2007ipython}, \texttt{Cython} \citep{behnel2010cython}, \texttt{Astropy} \citep{2013A&A...558A..33A, 2018AJ....156..123A}, \texttt{NumPy} \citep{van2011numpy}, \texttt{SciPy} \citep{scipy}, \texttt{scikit-Learn} \citep{pedregosa2011scikit}, \texttt{Matplotlib} \citep{hunter2007matplotlib}, HDBSCAN \citep{mcinnes2017hdbscan}.

\newpage

\bibliographystyle{aasjournal}

\bibliography{stream}



\appendix

\end{CJK*}
\end{document}